\documentclass[aps,prb,reprint,tightenlines,citeautoscript,showpacs,floatfix,superscriptaddress]{revtex4-1}
\usepackage{bm}
\usepackage{amssymb}
\usepackage{graphicx}
\usepackage{amsmath}
\usepackage{multirow}
\usepackage{float}

\begin{document}

\title{Nonlinear optics of graphene and other 2D materials in layered structures}

\author{J. L. Cheng}

\affiliation{The Guo China-US Photonics Laboratory (GPL), State Key Laboratory
of Applied Optics (SKLAO), Changchun Institute of Optics, Fine Mechanics
and Physics (CIOMP), Chinese Academy of Sciences (CAS), Changchun
130033, P R China.}

\affiliation{School of Physical Sciences, University of Chinese Academy of Sciences,
Beijing 100049, China}

\author{J. E. Sipe}

\affiliation{Department of Physics, University of Toronto, Toronto, Ontario, Canada}

\author{N. Vermeulen}

\affiliation{Brussels Photonics, Department of Applied Physics and Photonics,
Vrije Universiteit Brussel, Pleinlaan 2, 1050 Brussel, Belgium}

\author{C. Guo}
\email{chunlei.guo@rochester.edu}
\affiliation{The Guo China-US Photonics Laboratory (GPL), State Key Laboratory
of Applied Optics (SKLAO), Changchun Institute of Optics, Fine Mechanics
and Physics (CIOMP), Chinese Academy of Sciences (CAS), Changchun
130033, P R China.}
\affiliation{The Institute of Optics, University of Rochester, Rochester, NY 14627,
USA}

\date{\today}

\begin{abstract}
We present a theoretical framework for nonlinear optics of graphene
and other 2D materials in layered structures. We derive a key equation
to find the effective electric field and the sheet current density
in the 2D material for given incident light beams. Our approach takes into
account the effect of the surrounding environment and characterizes
its contribution as a structure factor. We apply our approach to two
experimental setups, and discuss the structure factors for several
nonlinear optical processes including second harmonic generation,
third harmonic generation, and parametric frequency conversion. Our
systematic study gives a strict extraction method for the nonlinear
coefficients, and provides new insights in how layered structures
influence the nonlinear signal observed from 2D materials. 
\end{abstract}

\maketitle

\section{Introduction}

Graphene possesses extraordinary optical properties \cite{Nat.Photon._4_611_2010_Bonaccorso},
including broadband absorption \cite{Science_320_1308_2008_Nair},
the existence of tightly confined plasmon modes \cite{ACSNano_8_1086_2014_Low,ACSPhotonics_1_135_2014_GarciadeAbajo,Proc.R.Soc.A_473_20170433_2017_Ooi},
and extremely strong optical nonlinearities \cite{Phys.Rep._535_101_2014_Glazov,Nat.Photon._10_227_2016_Sun,Adv.Mater._x_1705963_2018_Autere}.
Most of these properties strongly depend on the chemical potential,
which can be controlled chemically \cite{J.Phys.Condens.Matter_21_402001_2009_Pinto,J.Mater.Chem._21_3335_2011_Liu},
optically \cite{Nat.Photon._10_244_2016_Ni}, or with the use of an
external gate voltage \cite{Science_306_666_2004_Novoselov}. Since
the first experimental demonstration of strong parametric frequency
conversion (PFC) in graphene \cite{Phys.Rev.Lett._105_097401_2010_Hendry},
much attention has been paid to its optical nonlinearities. With the
integration of graphene onto photonic chips becoming
a mature technology \cite{Nat.Photon._4_611_2010_Bonaccorso,Nat.Photon._8_899_2014_Xia,IEEEJ.Sel.Top.QuantumElectron._22_347_2016_Vermeulen},
graphene is recognized as a potential resource for many photonic devices
that exploit optical nonlinearities, including saturable absorbers
\cite{Adv.Funct.Mater._19_3077_2009_Bao}, broadband optical modulators
\cite{Nature_474_64_2011_Liu,NanoLett._12_1482_2012_Liu,Appl.Phys.Lett._105_111110_2014_Ooi},
optical switches \cite{APLPhoton._1_46101_2016_Ooi}, and wavelength
converters \cite{IEEEJ.Sel.Top.QuantumElectron._22_347_2016_Vermeulen}.
For these applications to be developed, a full understanding of the
nonlinear radiation of graphene is necessary. In general, this is
determined by both the intrinsic optical nonlinearity of graphene,
and the effects of the surrounding environment. Neither is
sufficiently understood at this point.

The intrinsic optical nonlinearity of graphene has been studied
extensively. 
Various experiments with graphene in free-space configurations, on waveguides, and
on photonic crystals have been performed to investigate different
nonlinear phenomena, including PFC \cite{Phys.Rev.Lett._105_097401_2010_Hendry,Nat.Photon.___2018_Jiang},
third harmonic generation (THG) \cite{Phys.Rev.B_87_121406_2013_Kumar,Phys.Rev.X_3_021014_2013_Hong,ACSNano_7_8441_2013_Saeynaetjoki,Nat.Photon.___2018_Jiang,Nat.Nano._X_X_2018_Soavi},
second harmonic generation (SHG) \cite{Appl.Phys.Lett._95_261910_2009_Dean,Phys.Rev.B_82_125411_2010_Dean,Phys.Rev.B_85_121413_2012_Bykov,NanoLett._13_2104_2013_An,Phys.Rev.B_89_115310_2014_An,Appl.Phys.Lett._105_151605_2014_Lin},
Kerr effects (or self-phase modulation, SPM) and two photon absorption \cite{Opt.Express_18_4564_2010_Xing,NanoLett._11_5159_2011_Wu,NanoLett._11_2622_2011_Yang,Opt.Lett._37_1856_2012_Zhang,Nat.Photon._6_554_2012_Gu,Phys.Rev.Appl._6_044006_2016_Vermeulen,Opt.Lett._41_3281_2016_Dremetsika,NC_Nathalie},
and coherent current injection \cite{NanoLett._10_1293_2010_Sun,Phys.Rev.B_85_165427_2012_Sun}.
The extracted effective bulk third order susceptibilities
$\chi_{\text{eff}}^{(3)}$ 
are in the range of $10^{-13}-10^{-19}\text{~m}^{2}/\text{V}^{2}$,
while the values of the effective bulk second order susceptibilities
$\chi_{\text{eff}}^{(2)}$ are seldom extracted. Recent experiments 
have shown the unusual negative sign of graphene's nonlinear refractive
index \cite{Phys.Rev.Appl._6_044006_2016_Vermeulen,Opt.Lett._41_3281_2016_Dremetsika,Opt.Express_24_13033_2016_Demetriou}, the ability to tune the nonlinearity by varying the chemical
potential \cite{Nat.Photon.___2018_Jiang,Nat.Nano._X_X_2018_Soavi}, 
and the possibility of having quasi-exponentially growing
  SPM in  graphene on waveguides \cite{NC_Nathalie}. 
Most of the existing microscopic theories
\cite{NewJ.Phys._16_53014_2014_Cheng,*Corrigendum_NewJ.Phys._18_29501_2016_Cheng,Phys.Rev.B_91_235320_2015_Cheng,*Phys.Rev.B_93_39904_2016_Cheng,Phys.Rev.B_92_235307_2015_Cheng,Phys.Rev.A_94_023845_2016_Soh,Phys.Rev.B_93_161411_2016_Rostami,Phys.Rev.B_93_85403_2016_Mikhailov,Phys.Rev.B_94_195442_2016_Wang,arXiv:1609.06413,arXiv:1608.00877,arXiv:1605.06499,arXiv:1610.04854,Phys.Rev.B_97_115454_2018_Avetissian}
focus on the dependence
of the sheet conductivities $\sigma^{(n)}$ ($\propto-i\omega\chi^{(n)}_{\text{eff}}$)
on incident light frequencies, temperature, and doping levels. When
scattering is described within a phenomenological relaxation time
approximation, the calculated $\chi^{(3)}_{\text{eff}}$ are about
two orders of magnitude smaller than most values extracted from 
earlier experiments for lightly doped graphene \cite{NewJ.Phys._16_53014_2014_Cheng,*Corrigendum_NewJ.Phys._18_29501_2016_Cheng,Phys.Rev.B_91_235320_2015_Cheng,*Phys.Rev.B_93_39904_2016_Cheng,Phys.Rev.A_94_023845_2016_Soh}. 
Several effects have been considered that might bring theory in better
agreement with these experiments, including saturation of the optical nonlinearity
\cite{Phys.Rev.B_92_235307_2015_Cheng,arXiv:1605.06499}, the influence
of cascaded second order processes \cite{Nat.Phys._12_124_2015_Constant1,arXiv:1610.04854},
and novel plasmonic effects
\cite{Nat.Comm._5_5725_2014_Cox,arXiv:1610.04854}, but
none of them can
sufficiently enhance the calculated conductivities. Recent experiment
on light propagating in graphene-covered waveguides \cite{NC_Nathalie} has shown that
free-carrier refraction plays an important role, and that a treatment
beyond perturbation theory is required there. This may be necessary in
other excitation configurations as well. However, agreement with
perturbative calculations is satisfactory in recent free-space
experiments, where the doping level of graphene is tuned over a large
range \cite{Nat.Photon.___2018_Jiang,Nat.Nano._X_X_2018_Soavi}.

Besides the insufficient understanding of the physical mechanism for the
intrinsic optical nonlinearity, the
accurate extraction of the effective susceptibility itself from experimental data is also
an outstanding problem, especially considering the one-atom thickness.
Experimentally, two methods are widely used, both treating graphene
as a thin film \cite{Phys.Rev.Lett._105_097401_2010_Hendry} with
an effective thickness $d_{\text{gr}}\approx3.3$~\AA{}. One is simply
to compare the radiation signal to a film with known susceptibility,
and the other is to utilize the well-known results for a thin bulk
sample in the limit of an undepleted fundamental \cite{Phys.Rev.B_87_121406_2013_Kumar,boyd_nonlinearoptics}.
Neither strategy is based on a strict derivation of the response of
a monolayer sample, and the effective susceptibility extracted from
either method depends on the artificial thickness assigned to the
graphene layer. For usual bulk materials, the generated nonlinear
radiation can be worked out using coupled-mode theory \cite{boyd_nonlinearoptics},
in which a coherence length associated with the phase-matching condition
arises in the derivation of the equations for the propagating fields.
It has been pointed out that for weakly coupled 2D layers the
atomic spacing between the layers should guarantee coherent radiation
from the layers \cite{Adv.Mater._x_1705963_2018_Autere,LightSci.App._5_16131_2016_Zhao}.
Despite many investigations using standard software
\cite{Phys.Rev.B_87_121406_2013_Kumar,Opt.Lett._38_3550_2013_Vincenti,Phys.Rev.B_89_165139_2014_Vincenti},
and nonlinear boundary conditions
\cite{Appl.Phys.Lett._107_181104_2015_Savostianova,arXiv:1608.05975,Phys.Rev.A_94_013811_2016_Gorbach,Phys.Rev.A_87_013830_2013_Gorbach}, 
there is still no systematic treatment of the response of graphene,
taking into account its 2D nature, which could later be generalized
to treat multilayer samples. 

In this work we focus on how the surrounding environment affects the
nonlinear radiation of general 2D materials inside layered structures.
We model the current density in graphene as a ``current sheet''
described by a Dirac $\delta$-function, and derive a key equation
for graphene nonlinear optics, from which the effective electric fields
and current density inside 2D materials can be determined as a response
to incident laser beams. These results are easily combined with the
transfer matrix formalism, and are used to analyze the nonlinear fields
generated by a graphene sheet on multilayered structures. The output
fields $E_{\text{out}}$ are connected to incident fields $E_{\text{in}}$
in a form $E_{\text{out}}\propto{\sigma}^{(n)}\beta^{(n)}E_{\text{in}}^{n}$,
with the structure factor $\beta^{(n)}$ describing the environmental 
effects. In certain structures, the contribution from the structure
factor can be extremely large, providing a controllable way to enhance
the generated nonlinear signals \cite{Opt.Lett._38_3550_2013_Vincenti,Phys.Rev.B_89_165139_2014_Vincenti,Phys.Rev.Lett._112_55501_2014_Yao,Nat.Phys._12_124_2015_Constant1,Opt.Express_23_7809_2015_Tao}.
The equations we establish can be readily applied to the entire family
of 2D materials, including bi-layer graphene, functionalized graphene,
monolayer transition-metal dichalcogenides, black phosphorene, silicene,
stanene, and so on. We also discuss whether or not a cascaded second
order response can modify extracted effective third order nonlinear
response coefficients. 

We organize this paper as follows: In Sec.~\ref{sec:suspended2D}
we work out the equations for a suspended 2D layer, and obtain a consistent
set of equations for the second and third order nonlinear response;
in Sec.~\ref{sec:multi} we extend these results to a system with
2D materials embedded inside a multilayered structure; in Sec.~\ref{sec:graphene}
we apply these results to a graphene-covered multilayered structure.
We conclude in Sec.~\ref{sec:conclude}.

\section{Suspended 2D layer\label{sec:suspended2D}}
\begin{figure}[!htpb]
\centering \includegraphics[width=7cm]{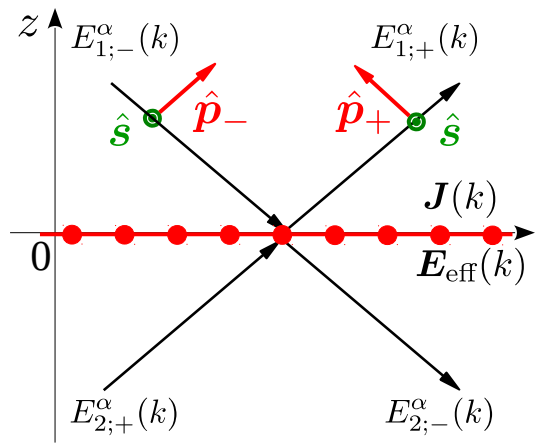}
\caption{(color online) A suspended 2D layer at position $z=0$. Here $E_{j;\lambda}^{\alpha}(k)$
are the amplitudes of the $\alpha$-polarized components of the electric
field propagating (or evanescent) in the $j$th region in the $\lambda$
direction; $\hat{\bm{s}}$ and $\hat{\bm{p}}_{\pm}$ are polarization
vectors, with a compact notation $k=(\bm{\kappa},\omega)$. The sketch
of the polarization vectors and the propagation directions is schematic,
since in general $\hat{\bm{p}}_{\pm}$can have complex Cartesian components
and the fields can be evanescent (See Appendix~\ref{app:notation}).
The 2D layer experiences an electric field $\bm{E}_{\text{eff}}(k)$
and possesses a sheet current $\bm{J}(k)$.}
\label{fig:2Doptic} 
\end{figure}
We first consider the optics of a suspended monolayer (or nearly-monolayer)
structure, which is centered at $z=0$, as shown in Fig.~\ref{fig:2Doptic}.
In the neighborhood of the monolayer the microscopic electric field
and current density have a very complicated spatial dependence, with
variations on an atomic scale, and their full description requires
a very careful treatment \cite{microscopic}. However, for the calculation
of radiation at optical frequencies with wavelengths much larger than
these dimensions, a ``$\delta$-model'' for the current density
can be introduced, 
\begin{equation}
\bm{J}(z;k)=\delta(z)\bm{J}(k)\,,\label{eq:2DPj}
\end{equation}
where $k=(\bm{\kappa},\omega)$ indicates the in-plane wave vector
$\bm{\kappa}$ and frequency $\omega$; for details of the notation
see Appendix~\ref{app:notation}. Here $\bm{J}(k)$ is a sheet current density.
We will see that it mostly appears in the form 
\begin{equation}
\vec{{\cal J}}(k)=\frac{\bm{J}(k)}{2c\epsilon_{0}}\,,
\end{equation}
which has the dimension of an electric field. 

Within this $\delta$-model all of space is split into three regions,
$z>0$, $z=0$, and $z<0$, and both the regions $z>0$ and $z<0$
are vacuum. The electric field in the latter two regions are described
by amplitudes $E_{j;\lambda}^{\alpha}(k)$, where $j=1,2$ is an index
identifying the region ($z>0$ and $z<0$ respectively), $\lambda=+$
($-$) stands for the upward (downward) propagating or evanescent
direction, and $\alpha=s$ ($p$) identifies the $s$ ($p$)-polarized
component. Using a Green function strategy \cite{J.Opt.Soc.Am.B_4_481_1987_Sipe}
(see Appendix~\ref{app:gfunc}), we can connect the fields in the
1st and 2nd regions by 

\begin{eqnarray}
E_{1;+}^{\alpha}(k) & = & E_{2;+}^{\alpha}(k)+{\cal F}_{+}^{\alpha}(k)\,,\label{eq:eu}\\
E_{2;-}^{\alpha}(k) & = & E_{1;-}^{\alpha}(k)+{\cal F}_{-}^{\alpha}(k)\,,\label{eq:ed}
\end{eqnarray}
with 
\begin{equation}
{\cal F}_{\lambda}^{\alpha}(k)=-\frac{\widetilde{\omega}}{w_{0}(k)}\vec{{\cal J}}(k)\cdot\hat{\bm{e}}_{0;\lambda}^{\alpha}(k)\,.\label{eq:calF2D}
\end{equation}
The vectors $\hat{\bm{e}}_{0;\lambda}^{\alpha}(k)$ are the
polarization vectors for the indicated 
polarizations $\alpha$ and propagation (or evanescent) directions
$\lambda$, which are defined in Appendix~\ref{app:notation}; $\widetilde{\omega}\equiv\omega/c$ and 
$w_{0}(k)=\sqrt{\widetilde{\omega}^{2}-\kappa^{2}}$. The fields in 
Eqs.~(\ref{eq:eu}) and (\ref{eq:ed}) include any incident fields that
are present, from sources or media 
above (amplitude $E_{1;-}^{\alpha}(k)$) or below (amplitude $E_{2;+}^{\alpha}(k)$)
the monolayer. 

To determine $\vec{{\cal J}}(k)$ and thus find the total fields above
and below the monolayer one needs to specify the response of the medium
to the full field, including that from the monolayer itself. Dealing with
the $\boldsymbol{\hat{z}}$ component of the response is here particularly
difficult, because the
$\boldsymbol{\hat{z}}$ component of the field would vary strongly
over that microscopic thickness. Largely this strongly varying field would
be included in a standard calculation of the response as done in condensed
matter physics, and thus would be included in such a calculation of
the linear or nonlinear response coefficients; we need to identify
an effective driving field $\bm{E}_{\text{eff}}(k)$ that would lead
to a calculation of these response coefficients. We follow the standard
approach (see, e.g. \cite{Phys.Rev.A_94_013811_2016_Gorbach}) and
take $\bm{E}_{\text{eff}}(k)$ to be the average of the fields above
and below the 2D layer, 
\begin{equation}
\bm{E}_{\text{eff}}(k)=\frac{1}{2}\sum_{\lambda\alpha}[E_{1;\lambda}^{\alpha}(k)+E_{2;\lambda}^{\alpha}(k)]\hat{\bm{e}}_{0;\lambda}^{\alpha}(k)\,.\label{eq:totE2D}
\end{equation}
The task of determining the dependence of $\vec{{\cal J}}(k)$ on
$\bm{E}_{\text{eff}}$(k) is then assigned to condensed matter physics;
it has been widely studied both perturbatively \cite{NewJ.Phys._16_53014_2014_Cheng,*Corrigendum_NewJ.Phys._18_29501_2016_Cheng,Phys.Rev.B_91_235320_2015_Cheng,*Phys.Rev.B_93_39904_2016_Cheng,Phys.Rev.B_93_85403_2016_Mikhailov,Phys.Rev.B_93_161411_2016_Rostami,arXiv:1610.04854}
and numerically \cite{Phys.Rev.B_92_235307_2015_Cheng,Phys.Rev.B_97_115454_2018_Avetissian}.
Other strategies and their associated definitions of $\bm{E}_{\text{eff}}(k)$
could be considered, but would just shift how much of the problem
was relegated to the task of determining $\vec{{\cal J}}[k;\bm{E}_{\text{eff}}(k)],$
and how much was relegated to the task of constructing $\bm{E}_{\text{eff}}(k)$
from the incident field and the field from the microscopic current
density itself. The advantage of the use of Eq.~(\ref{eq:totE2D}) is
that it is a simple expression but still completely contains the radiation
reaction field due to the current sheet, so that the effect of $\bm{E}_{\text{eff}}(k)$
on the sheet includes the consequences of its radiation carrying energy
away; energy conservation is thus respected.

\subsection{Effective fields and sheet current density inside the 2D layer \label{sec:o2d}}

Once the functional $\vec{{\cal J}}[k;\bm{E}_{\text{eff}}]$ is known,
Eqs.~(\ref{eq:eu}), (\ref{eq:ed}), and (\ref{eq:totE2D}) are closed.
Substituting Eqs.~(\ref{eq:eu}) and (\ref{eq:ed}) into Eq.~(\ref{eq:totE2D})
we express the effective fields from $\vec{{\cal J}}(k)$ as 
\begin{equation}
\bm{E}_{\text{eff}}(k)=\bm{E}_{\text{2D}}^{0}(k)+{\cal Q}^{0}(k)\vec{{\cal J}}(k)\,.\label{eq:radE}
\end{equation}
Here $\bm{E}_{\text{2D}}^{0}(k)=\sum_{\alpha}E_{1;-}^{\alpha}(k)\hat{\bm{e}}_{0;-}^{\alpha}(k)+\sum_{\alpha}E_{2;+}^{\alpha}(k)\hat{\bm{e}}_{0;+}^{\alpha}(k)$
is the incident field at the 2D layer (the field without the
presence of the 2D layer
), and ${\cal Q}^{0}(k)=-\frac{\widetilde{\omega}}{w_{0}}\hat{\bm{s}}(k)\hat{\bm{s}}(k)-\frac{w_{0}}{\widetilde{\omega}}\hat{\bm{\kappa}}\hat{\bm{\kappa}}-\frac{\kappa^{2}}{w_{0}\widetilde{\omega}}\hat{\bm{z}}\hat{\bm{z}}$
is a radiation coefficient tensor matrix associated with the 2D
layer. The second term at the right 
hand side of Eq.~(\ref{eq:radE}) gives the onsite fields induced
by the sheet current. Equation~(\ref{eq:radE}) is our key result;
what is also needed to solve for the radiated fields is the identification
of the functional $\vec{{\cal J}}[k;\bm{E}_{\text{eff}}]$ from a
response theory in condensed matter physics. Using $\vec{{\cal J}}[k;\bm{E}_{\text{eff}}]$
together with Eq.~(\ref{eq:radE}) then leads to a self-consistent solution
for the fields and the sheet current.

For weak electric fields, the sheet current density can be written
into a power expansion of the effective fields. Up to the third order
terms, it is 
\begin{eqnarray}
\vec{{\cal J}}(k) & = & \eta^{(1)}(k)\bm{E}_{\text{eff}}(k)+\vec{{\cal J}}_{\text{nl}}(k)\,,\label{eq:caljresp}\\
\vec{{\cal J}}_{\text{nl}}(k) & = & \int\frac{dk_{1}}{(2\pi)^{3}}\eta^{(2)}(k_{1},k-k_{1})\bm{E}_{\text{eff}}(k_{1})\bm{E}_{\text{eff}}(k-k_{1})\nonumber \\
 &  & +\int\frac{dk_{1}dk_{2}}{(2\pi)^{6}}\eta^{(3)}(k_{1},k_{2},k-k_{1}-k_{2})\nonumber \\
 &  & \quad\times\bm{E}_{\text{eff}}(k_{1})\bm{E}_{\text{eff}}(k_{2})\bm{E}_{\text{eff}}(k-k_{1}-k_{2})\,.\label{eq:jnlesh}
\end{eqnarray}
Here the tensors $\eta^{(1)}(k)=\sigma^{(1)}(k)/(2c\epsilon_{0})$,
$\eta^{(2)}(k_{1},k_{2})=\sigma^{(2)}(k_{1},k_{2})/(2c\epsilon_{0})$,
and $\eta^{(3)}(k_{1},k_{2},k_{3})=\sigma^{(3)}(k_{1},k_{2},k_{3})/(2c\epsilon_{0})$
are associated respectively with the linear, second order, and third
order conductivities.

The linear conductivity $\sigma^{(1)}(k)$ has been widely discussed
\cite{J.Phys.Soc.Jpn._71_1318_2002_Ando,Phys.Rev.B_73_245411_2006_Gusynin}.
In our calculations it is a good approximation to set $\sigma^{(1)}(k)\approx\sigma^{(1)}(\bm{0},\omega)\equiv\sigma^{(1)}(\omega)$
because in the microscopic calculation of the conductivity the light
wave vector is much smaller than the involved electron wave vectors,
and is usually ignored. Furthermore, for most 2D materials, the linear
conductivity tensor only has nonzero components $\sigma^{(1);xx}=\sigma^{(1);yy}=\sigma_{\parallel}^{(1)}(\omega)$
and $\sigma^{(1);zz}=\sigma_{\perp}^{(1)}(\omega)$. Although $\sigma_{\perp}^{(1)}(\omega)$
is usually ignored, its value is not zero and it is important for
certain applications \cite{Proc.R.Soc.A_473_20170433_2017_Ooi}; as
well, including it would be useful in characterizing samples. As with
$\sigma^{(1)}(k)$, it is also a good approximation to neglect the
dependence of $\sigma^{(3)}$ on the wave vector and set all $\bm{\kappa}=0$.
While for $\sigma^{(2)}$ this would hold if the material lacked inversion
symmetry, in the presence of such symmetry (such as in graphene),
at least the dependence on $\bm{\kappa}$ to first order must be kept,
which corresponds to keeping electric-quadrupole-like and magnetic-dipole-like
contributions to the second order nonlinear response\cite{arXiv:1609.06413,Phys.Rev.B_94_195442_2016_Wang,arXiv:1610.04854}.

Usually the nonlinear term in Eq.~(\ref{eq:caljresp}) is much smaller
than the linear term, so it is convenient to isolate the nonlinear response.
From Eqs.~(\ref{eq:radE}) and (\ref{eq:caljresp}) the effective
field at the 2D layer is formally given by \begin{subequations} 
\begin{equation}
\bm{E}_{\text{eff}}(k)=\left[\hat{I}-{\cal Q}^{0}(k)\eta^{(1)}(k)\right]^{-1}\left[\bm{E}_{\text{2D}}^{0}(k)+{\cal Q}^{0}(k)\vec{{\cal J}}_{\text{nl}}(k)\right]\,,\label{eq:radE2}
\end{equation}
and the sheet current density is 
\begin{equation}
\vec{{\cal J}}(k)=\left[\hat{I}-\eta^{(1)}(k){\cal Q}^{0}(k)\right]^{-1}\left[\eta^{(1)}(k)\bm{E}_{\text{2D}}^{0}(k)+\vec{{\cal J}}_{\text{nl}}(k)\right]\,,\label{eq:suspj}
\end{equation}
\end{subequations} with $\hat{I}=\hat{\bm s}(k)\hat{\bm s}(k)+\hat{\bm
  k}\hat{\bm k}+\hat{\bm z}\hat{\bm z}$ being the unit dyadic.

In this paper, we focus on the perturbative solutions of
Eqs.~(\ref{eq:radE2}) and (\ref{eq:jnlesh}). With respect to the
effective fields, we can expand $\bm{E}_{\text{eff}}(k)=\bm{E}_{\text{eff}}^{(1)}(k)+\bm{E}_{\text{eff}}^{(2)}(k)+\cdots$
and $\vec{{\cal J}}_{\text{nl}}(k)=\vec{{\cal J}}_{\text{nl}}^{(2)}(k)+\vec{{\cal J}}_{\text{nl}}^{(3)}(k)+\cdots$,
with \begin{subequations} 
\begin{align}
\bm{E}_{\text{eff}}^{(1)}(k) & =\left[\hat{I}-{\cal Q}^{0}(k)\eta^{(1)}(k)\right]^{-1}\bm{E}_{\text{2D}}^{0}(k)\,,\label{eq:Eeff1}\\
\vec{{\cal J}}_{\text{nl}}^{(2)}(k) & =\int\frac{dk_{1}}{(2\pi)^{3}}\eta^{(2)}(k_{1},k-k_{1})\nonumber \\
 & \times\bm{E}_{\text{eff}}^{(1)}(k_{1})\bm{E}_{\text{eff}}^{(1)}(k-k_{1})\,,\label{eq:caljnl2}
\end{align}
and 
\begin{align}
\bm{E}_{\text{eff}}^{(2)}(k) & =\left\{ [{\cal Q}^{0}(k)]^{-1}-\eta^{(1)}(k)\right\} ^{-1}\vec{{\cal J}}_{\text{nl}}^{(2)}(k)\,,\label{eq:Eeff2}\\
\vec{{\cal J}}_{\text{nl}}^{(3)}(k) & =\int\frac{dk_{1}dk_{2}}{(2\pi)^{6}}\eta^{(3)}(k_{1},k_{2},k-k_{1}-k_{2})\nonumber \\
 & \hspace{0.5cm}\times\bm{E}_{\text{eff}}^{(1)}(k_{1})\bm{E}_{\text{eff}}^{(1)}(k_{2})\bm{E}_{\text{eff}}^{(1)}(k-k_{1}-k_{2})\nonumber \\
 & +\int\frac{dk_{1}}{(2\pi)^{3}}\eta^{(2)}(k_{1},k-k_{1})[\bm{E}_{\text{eff}}^{(2)}(k_{1})\bm{E}_{\text{eff}}^{(1)}(k-k_{1})\nonumber \\
 & \hspace{0.5cm}+\bm{E}_{\text{eff}}^{(1)}(k_{1})\bm{E}_{\text{eff}}^{(2)}(k-k_{1})]\,.\label{eq:caljnl3}
\end{align}
\end{subequations} The third order term $\vec{{\cal J}}_{\text{nl}}^{(3)}(k)$
has two contributions: One is from the direct third order processes,
which is widely discussed in literature; the other is from the cascaded second order
processes. The results of a carefully designed experiment
\cite{Nat.Phys._12_124_2015_Constant1} suggest that cascaded
second order processes could lead to a significant change in the reflection
spectrum due to the generation of surface plasmons.

\subsection{Outgoing fields}
Substituting Eq.~(\ref{eq:suspj}) into Eqs.~(\ref{eq:eu}) and
(\ref{eq:ed}), we get the outgoing fields as \begin{subequations}
\begin{align}
E_{1;+}^{\alpha}(k) & =r^{\alpha}(k)E_{1;-}^{\alpha}(k)+t^{\alpha}(k)E_{2;+}^{\alpha}(k)\nonumber \\
 & +\frac{1+t^{\alpha}(k)}{2}{\cal G}_{\text{2D};+}^{\alpha}(k)+\frac{r^{\alpha}(k)}{2}\mathcal{G}_{\text{2D};-}^{\alpha}(k)\,,\label{eq:outfa}\\
E_{2;-}^{\alpha}(k) & =r^{\alpha}(k)E_{2;+}^{\alpha}(k)+t^{\alpha}(k)E_{1;-}^{\alpha}(k)\nonumber \\
 & +\frac{1+t^{\alpha}(k)}{2}{\cal G}_{\text{2D};-}^{\alpha}(k)+\frac{r^{\alpha}(k)}{2}\mathcal{G}_{\text{2D};+}^{\alpha}(k)\,.\label{eq:outfb}
\end{align}
\end{subequations} Here $r^{\alpha}(k)$ and $t^{\alpha}(k)$ are
Fresnel coefficients of a suspended 2D layer, and are given by \begin{subequations}
\begin{align}
t^{s}(k) & =\frac{1}{1+\frac{\widetilde{\omega}}{w_{0}(k)}\eta_{\parallel}^{(1)}(\omega)}\,,\quad r^{s}(k)=t^{s}(k)-1\,,\label{eq:fresnels}\\
r^{p}(k) & =-\frac{1}{1+\frac{w_{0}(k)}{\widetilde{\omega}}\eta_{\parallel}^{(1)}(\omega)}+\frac{1}{1+\frac{\kappa^{2}}{\widetilde{\omega}w_{0}(k)}\eta_{\perp}^{(1)}(\omega)}\,,\label{eq:fresnelpr}\\
t^{p}(k) & =\frac{1}{1+\frac{w_{0}(k)}{\widetilde{\omega}}\eta_{\parallel}^{(1)}(\omega)}-1+\frac{1}{1+\frac{\kappa^{2}}{\widetilde{\omega}w_{0}(k)}\eta_{\perp}^{(1)}(\omega)}\,.\label{eq:fresnelpt}
\end{align}
\end{subequations} The radiation term from the nonlinear current
is 
\begin{equation}
{\cal G}_{\text{2D};\lambda}^{\alpha}(k)=-\frac{\widetilde{\omega}}{w_{0}(k)}\vec{{\cal J}}_{\text{nl}}(k)\cdot\hat{\bm{e}}_{0,\lambda}^{\alpha}(k)\,.\label{eq:nong}
\end{equation}
Note that in the absence of a nonlinear current the results in
Eqs.~(\ref{eq:outfa}) and (\ref{eq:outfb})
in terms of the Fresnel coefficients are just what one would expect.

For use in the next section it is convenient to write these results
in a transfer matrix formalism. We can rewrite Eqs.~(\ref{eq:outfa})
and (\ref{eq:outfb}) to give 
\begin{eqnarray}
\begin{pmatrix}{E}_{1;+}^{\alpha}(k)\\
{E}_{1;-}^{\alpha}(k)
\end{pmatrix} & = & {\cal M}_{\text{2D}}^{\alpha}(k)\begin{pmatrix}{E}_{2;+}^{\alpha}(k)\\
{E}_{2;-}^{\alpha}(k)
\end{pmatrix}\nonumber \\
 & + & \frac{{\cal M}_{\text{2D}}^{\alpha}(k)+1}{2}\begin{pmatrix}{\cal G}_{\text{2D};+}^{\alpha}(k)\\
-{\cal G}_{\text{2D};-}^{\alpha}(k)
\end{pmatrix}\,,\label{eq:lo}
\end{eqnarray}
where ${\cal M}_{\text{2D}}^{\alpha}(k)$ is a transfer matrix for
the 2D material layer, and it is given from the Fresnel coefficients
as 
\begin{equation}
{\cal M}_{\text{2D}}^{\alpha}(k)=\frac{1}{t^{\alpha}(k)}\begin{pmatrix}[t^{\alpha}(k)]^{2}-[r^{\alpha}(k)]^{2} & r^{\alpha}(k)\\
-r^{\alpha}(k) & 1
\end{pmatrix}\,,
\end{equation}

For a $s$- or $p$-polarized light beam incident from above, the
absorption induced by this suspended layer is given by $1-|r^{\alpha}|^{2}-|t^{\alpha}|^{2}$.
For $s$-polarized incident light, the absorption is ${2\text{Re}\left[\frac{\widetilde{\omega}}{w_{0}(k)}\eta_{\parallel}^{(1)}(\omega)\right]}/{\left|1+\frac{\widetilde{\omega}}{w_{0}(k)}\eta_{\parallel}^{(1)}(\omega)\right|^{2}}$.
At normal incidence, where the difference between $s-$ and $p-$polarized
light vanishes, it is $2\eta_{0}^{(1)}/(1+\eta_{0}^{(1)})^{2}\sim2.24\%$
for intrinsic graphene with
$\eta_{0}^{(1)}=\sigma_{0}/(2c\epsilon_{0})$ and the universal
conductivity $\sigma_{0}=e^2/(4\hbar)$,
which agrees with the experimental value \cite{Science_320_1308_2008_Nair}.

\section{2D layer inside multilayers\label{sec:multi}}

\begin{figure}[!htpb]
\centering \includegraphics[width=8cm]{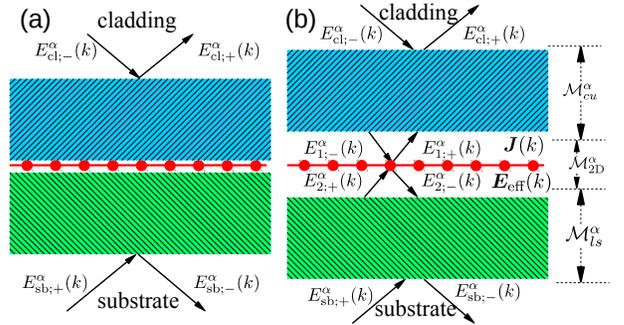}
\caption{(color online) (a) 2D layer (red in the middle) within a multilayered
structure. The incident fields are $E_{\text{cl};-}^{\alpha}(k)$
from cladding and $E_{\text{sb};+}^{\alpha}(k)$ from substrate, while
the outgoing fields are $E_{\text{cl};+}^{\alpha}(k)$ and $E_{\text{sb};-}^{\alpha}(k)$.
(b) An equivalent structure of (a) by formally inserting a  vacuum
region around the 2D layer, similar to Fig.~\ref{fig:2Doptic}. The
transfer matrix ${\cal M}_{cu}^{\alpha}(k)$ is for the region from
the cladding to the upper surface of the region containing the 2D layer,
and ${\cal M}_{ls}^{\alpha}(k)$ is for the region from the lower
surface of the region containing the 2D layer to the substrate.}
\label{fig:2Doptic1} 
\end{figure}

In this section we consider the effects of the surrounding environment
on the nonlinear optics of a 2D material, with a structure shown in
Fig.~\ref{fig:2Doptic1} (a). The incident fields are $E_{\text{cl};-}^{\alpha}(k)$
from the cladding and $E_{\text{sb};+}^{\alpha}(k)$ from the substrate,
while the outgoing fields are $E_{\text{cl};+}^{\alpha}(k)$ and $E_{\text{sb};-}^{\alpha}(k)$.
We only consider the nonlinear optics from the 2D layer; for other
layers, nonlinear radiation can be obtained from the coupled mode
theory \cite{boyd_nonlinearoptics} or, in an undepleted pump approximation,
by the bulk medium version of these equations \cite{J.Opt.Soc.Am.B_4_481_1987_Sipe}
. To utilize the transfer matrix formalism, we slightly deform the
structure into Fig.~\ref{fig:2Doptic1} (b), anticipating that we
will treat the thickness of the vacuum region containing the 2D layer
as vanishingly small. Now the expressions in Sec.~\ref{sec:suspended2D}
can be straightforwardly extended. We denote the transfer matrix for
the multilayered structure from the cladding layer to the upper surface
of the 2D layer region as ${\cal M}_{cu}^{\alpha}(k)$, and the transfer
matrix for the multilayered structure from the lower surface of the
2D layer region to the substrate as ${\cal M}_{ls}^{\alpha}(k)$.
Then we have \begin{subequations} 
\begin{align}
\begin{pmatrix}E_{\text{cl};+}^{\alpha}(k)\\
E_{\text{cl};-}^{\alpha}(k)
\end{pmatrix} & ={\cal M}_{cu}^{\alpha}(k)\begin{pmatrix}E_{1;+}^{\alpha}(k)\\
E_{1;-}^{\alpha}(k)
\end{pmatrix}\,,\label{eq:TMu}\\
\begin{pmatrix}E_{2;+}^{\alpha}(k)\\
E_{2;-}^{\alpha}(k)
\end{pmatrix} & ={\cal M}_{ls}^{\alpha}(k)\begin{pmatrix}E_{\text{sb};+}^{\alpha}(k)\\
E_{\text{sb};-}^{\alpha}(k)
\end{pmatrix}\,,\label{eq:TMd}
\end{align}
\end{subequations} Eliminating $E_{j;\pm}^{\alpha}(k)$ from Eqs.~(\ref{eq:lo}),
(\ref{eq:TMu}), and (\ref{eq:TMd}), the fields at the cladding connect
with the fields at the substrate by a transfer matrix formalism as
\begin{equation}
\begin{pmatrix}E_{\text{cl};+}^{\alpha}(k)\\
E_{\text{cl};-}^{\alpha}(k)
\end{pmatrix}={\cal M}_{cs}^{\alpha}(k)\begin{pmatrix}E_{\text{sb};+}^{\alpha}(k)\\
E_{\text{sb};-}^{\alpha}(k)
\end{pmatrix}+{\cal M}_{\text{nl}}^{\alpha}(k)\begin{pmatrix}{\cal G}_{\text{2D};+}^{\alpha}(k)\\
-{\cal G}_{\text{2D};-}^{\alpha}(k)
\end{pmatrix}\,,
\end{equation}
with the total transfer matrix 
\begin{equation}
{\cal M}_{cs}^{\alpha}(k)={\cal M}_{cu}^{\alpha}(k){\cal M}_{\text{2D}}^{\alpha}(k){\cal M}_{ls}^{\alpha}(k)\,,
\end{equation}
and a transfer matrix for the nonlinear radiation 
\begin{equation}
{\cal M}_{\text{nl}}^{\alpha}(k)={\cal M}_{cu}^{\alpha}(k)\frac{{\cal M}_{\text{2D}}^{\alpha}(k)+1}{2}\,.
\end{equation}
Converting to a scattering matrix form, we obtain the outgoing fields
as 
\begin{eqnarray}
\begin{pmatrix}E_{\text{cl};+}^{\alpha}(k)\\
E_{\text{sb};-}^{\alpha}(k)
\end{pmatrix} & = & \begin{pmatrix}R_{cs}^{\alpha}(k) & T_{cs}^{\alpha}(k)\\
T_{sc}^{\alpha}(k) & R_{sc}^{\alpha}(k)
\end{pmatrix}\begin{pmatrix}E_{\text{cl};-}^{\alpha}(k)\\
E_{\text{sb};+}^{\alpha}(k)
\end{pmatrix}\nonumber \\
 & + & \begin{pmatrix}1 & -R_{cs}^{\alpha}(k)\\
0 & -T_{sc}^{\alpha}(k)
\end{pmatrix}{\cal M}_{\text{nl}}^{\alpha}(k)\begin{pmatrix}{\cal G}_{\text{2D};+}^{\alpha}(k)\\
-{\cal G}_{\text{2D};-}^{\alpha}(k)
\end{pmatrix}\,,\quad\quad\label{eq:outE0}
\end{eqnarray}
where the Fresnel coefficients appearing here are associated with
the total transfer matrix ${\cal M}_{cs}^{\alpha}(k)$ by 
\begin{equation}
{\cal M}_{cs}^{\alpha}(k)=\frac{1}{T_{sc}^{\alpha}(k)}\begin{pmatrix}{T}_{cs}^{\alpha}(k){T}_{sc}^{\alpha}(k)-{R}_{cs}^{\alpha}(k){R}_{sc}^{\alpha}(k) & {R}_{cs}^{\alpha}(k)\\
-{R}_{sc}^{\alpha}(k) & 1
\end{pmatrix}\,.\label{eq:RtoM}
\end{equation}

In the absence of the nonlinear terms the results from Eq.~(\ref{eq:outE0})
for $E_{\text{cl};+}^{\alpha}(k)$ and $E_{\text{sb};-}^{\alpha}(k)$
in terms of the Fresnel coefficients are just what one would expect. For the nonlinear radiation, the key quantity
is still ${\cal G}_{\text{2D};\lambda}^{\alpha}(k)$, or equivalently
the nonlinear current $\vec{{\cal J}}_{\text{nl}}(k)$, which strongly
depend on the total field $\bm{E}_{\text{eff}}(k)$ at the 2D layer.
Again, similar to the strategy for a suspended 2D layer, it can be
found from Eqs.~(\ref{eq:eu}), (\ref{eq:ed}), (\ref{eq:totE2D}),
(\ref{eq:TMu}), and (\ref{eq:TMd}) as 
\begin{equation}
\bm{E}_{\text{eff}}(k)=\bm{E}_{\text{2D}}(k)+{\cal Q}(k)\vec{{\cal J}}(k)\,,\label{eq:radE1}
\end{equation}
where the incident fields $\bm{E}_{\text{2D}}(k)$ at the 2D layer,
and the radiation coefficients ${\cal Q}(k)$ are given in Appendix~\ref{app:factor}.
Equation~(\ref{eq:radE1}) has the same structure as Eq.~(\ref{eq:radE}),
but each term is modified to include the effects from the environment,
\textit{i.e.}, other layers. All the results from Eq.~(\ref{eq:radE2})
to Eq.~(\ref{eq:caljnl3}) can be simply transferred here after replacing
$\bm{E}_{\text{2D}}^{0}(k)$ and ${\cal Q}^{0}(k)$ by $\bm{E}_{2D}(k)$
and ${\cal Q}(k)$, respectively. For example, comparing to Eq.~(\ref{eq:radE2}), the effective field
depends on the nonlinear current according to 
\begin{eqnarray}
\bm{E}_{\text{eff}}(k)=\left[\hat{I}-{\cal Q}(k)\eta^{(1)}(k)\right]^{-1}\left[\bm{E}_{\text{2D}}(k)+{\cal Q}(k)\vec{{\cal J}}_{\text{nl}}(k)\right]\,.\quad\quad\label{eq:fullEeff}
\end{eqnarray}


\section{Results for a graphene monolayer\label{sec:graphene}}
\begin{figure}[!htp]
\centering \includegraphics[width=7cm]{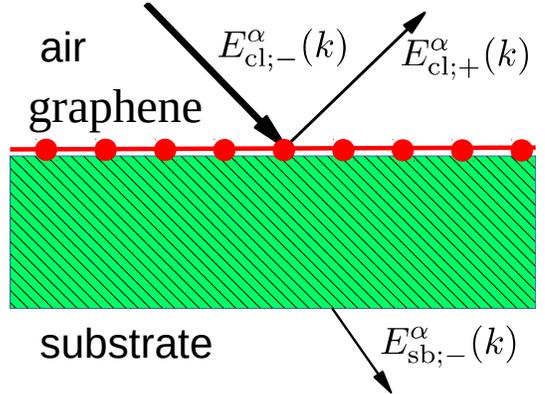}
\caption{(color online) Illustration of a graphene-covered multilayered structure.}
\label{fig:graphene} 
\end{figure}
We illustrate our approach by considering a graphene-covered multilayered
structure shown in Fig.~\ref{fig:graphene}. We assume there is incident
light only from the cladding layer, so ${E}_{\text{sb};+}^{\alpha}(k)=0$,
and we are interested in the output field ${E}_{\text{cl};+}^{\alpha}(k)$.
We ignore any nonlinear sources in the multilayer, cladding, and substrate.
In the widely accepted treatment for graphene the optical response
in the $z$ direction is ignored, so we put $\eta_{\perp}^{(1)}(\omega)\rightarrow0$,
and assume as well that the nonlinear current has only in-plane components
and responds only to the in-plane electric field components. We then
require only the in-plane components of the effective fields, which
we write as $\bm{E}_{\text{eff};\parallel}(k)=E_{\text{eff};s}(k)\hat{\bm{s}}(k)+E_{\text{eff};\kappa}(k)\hat{\bm{\kappa}}$,
and similarly will have only an in-plane nonlinear current $\vec{{\cal J}}_{\text{nl};\parallel}(k)$.
With the $z$-component of the current sheet thus ignored and the
$z$-component of the effective field not required, Eq.~(\ref{eq:fullEeff})
is greatly simplified to give 
\begin{align}
E_{\text{eff};j}(k) & =\frac{E_{\text{2D};j}(k)+{\cal Q}_{jj}(k){\cal J}_{\text{nl};j}(k)}{1-{\cal Q}_{jj}(k)\eta_{\parallel}^{(1)}(k)}\,,
\end{align}
for $j=s$ or $\kappa$. Using Eq.~(\ref{eq:e2dsgh}) and
(\ref{eq:e2dkgh}) for $E_{\text{2D};j}(k)$,
the in-plane effective electric fields are found to be 
\begin{eqnarray}
E_{\text{eff};s}(k) & = & [1+R_{cs}^{s}(k)]\left[E_{\text{cl};-}^{s}(k)-\frac{\widetilde{\omega}}{w_{0}(k)}{\cal J}_{\text{nl};s}(k)\right]\,,\quad\quad\\
E_{\text{eff};\kappa}(k) & = & [1-R_{cs}^{p}(k)]\left[E_{\text{cl};-}^{p}(k)-{\cal J}_{\text{nl};\kappa}(k)\right]\frac{w_{0}(k)}{\widetilde{\omega}}\,.\quad\quad,
\end{eqnarray}
and from the transfer matrices the total reflectivity $R_{cs}^{\alpha}(k)$
is found to be
\begin{align}
R_{cs}^{\alpha}(k) & =r^{\alpha}(k)+\frac{t^{\alpha}(k)R_{ls}^{\alpha}(k)t^{\alpha}(k)}{1-R_{ls}^{\alpha}(k)r^{\alpha}(k)}\notag\\
 & =\frac{\tau^{\alpha}}{\dfrac{1}{1+\tau^{\alpha}R_{ls}^{\alpha}(k)}+\left[\dfrac{\widetilde{\omega}}{w_{0}(k)}\right]^{\tau^{\alpha}}\eta_{\parallel}^{(1)}(\omega)}-\tau^{\alpha}\,.
\end{align}
Here $R_{ls}^{\alpha}(k)$ is the reflection coefficient without the
presence of graphene, $\tau^{s}=1$, and $\tau^{p}=-1$; in the second
expression the result is displayed explicitly in terms of the linear
conductivity of the 2D layer. Finally, from Eq.~(\ref{eq:outE0})
the output fields can then be written into terms of the nonlinear
current as 
\begin{align}
{E}_{\text{cl},+}^{\alpha}(k) & =R_{cs}^{\alpha}(k){E}_{\text{cl},-}^{\alpha}(k)\nonumber \\
 & -[1+\tau^{\alpha}R_{cs}^{\alpha}(k)]\frac{\widetilde{\omega}}{w_{0}(k)}\vec{{\cal J}}_{\text{nl};\parallel}(k)\cdot\hat{\bm{e}}_{0;+}^{\alpha}\,.\label{eq:outE}
\end{align}

According to earlier results
\cite{Phys.Rev.B_91_235320_2015_Cheng,*Phys.Rev.B_93_39904_2016_Cheng,
  arXiv:1609.06413},
the nonlinear currents up to the third order terms are 
\begin{widetext}
{\allowdisplaybreaks 
\begin{align}
\vec{{\cal J}}_{\text{nl};\parallel}(k) & =2\int\frac{dk_{1}}{(2\pi)^{3}}\Bigg\{{\cal S}_{M}^{xxyy}(\omega_{1},\omega_{a})[\bm{\kappa}_{1}\times\bm{E}_{\text{eff};\parallel}(k_{1})]\times\bm{E}_{\text{eff};\parallel}(k_{a})+{\cal S}_{Q}^{xxyy}(\omega_{1},\omega_{a})[\bm{E}_{\text{eff};\parallel}(k_{1})\bm{\kappa}_{1}\cdot\bm{E}_{\text{eff};\parallel}(k_{a})\nonumber \\
 & \hspace{1cm}+\bm{\kappa}_{1}\bm{E}_{\text{eff};\parallel}(k_{1})\cdot\bm{E}_{\text{eff};\parallel}(k_{a})]+{\cal S}_{Q}^{xyxy}(\omega_{1},\omega_{a})\bm{\kappa}_{1}\cdot\bm{E}_{\text{eff};\parallel}(k_{1})\bm{E}_{\text{eff};\parallel}(k_{a})\Bigg\}\nonumber \\
 & +3\int\frac{dk_{1}dk_{2}}{(2\pi)^{6}}\eta^{(3);xxyy}(\omega_{1},\omega_{2},\omega_{b})\bm{E}_{\text{eff};\parallel}(k_{1})\bm{E}_{\text{eff};\parallel}(k_{2})\cdot\bm{E}_{\text{eff};\parallel}(k_{b})\,,\label{eq:jnl3}
\end{align}
} 
\end{widetext}
with $k_{a}=k-k_{1}$ and $k_{b}=k-k_{1}-k_{2}$. For the second order
response, ${\cal S}_{Q}^{xxyy}$ and ${\cal S}_{Q}^{xyxy}$ are response
coefficients associated with the quadrupole-like contribution, and ${\cal S}_{M}^{xxyy}$
is a response coefficient associated with the magnetic dipole-like
contribution; they are related to the parameters $S_{Q/M}^{dabc}$
defined earlier \cite{arXiv:1609.06413} by ${\cal S}=S/(2c\epsilon_{0})$.
For later use, the components along the $x$-direction of the conductivity
tensors are given by \begin{subequations} 
\begin{eqnarray}
\eta_{2}(k_{1},k_{2}) & \equiv & \eta^{(2);xxx}(k_{1},k_{2})\nonumber \\
 & = & {\cal S}^{xxxx}(\omega_{1},\omega_{2})\kappa_{1}^{c}+{\cal S}^{xxxx}(\omega_{2},\omega_{1})\kappa_{2}^{c}\,,\\
\eta_{3}(\omega_{1},\omega_{2},\omega_{3}) & \equiv & \eta^{(3);xxxx}(\omega_{1},\omega_{2},\omega_{3})\nonumber \\
 & = & \eta^{(3);xxyy}(\omega_{1},\omega_{2},\omega_{3})+\eta^{(3);xxyy}(\omega_{2},\omega_{3},\omega_{1})\nonumber \\
 &  & +\eta^{(3);xxyy}(\omega_{3},\omega_{1},\omega_{2})\,,
\end{eqnarray}
\end{subequations} with ${\cal S}^{xxxx}(\omega_{1},\omega_{2})=2{\cal S}_{Q}^{xxyy}(\omega_{1},\omega_{2})+{\cal S}_{Q}^{xyxy}(\omega_{1},\omega_{2})$.

We note that the argument $k=(\bm{\kappa},\omega)$ of ${E}_{\text{cl};-}^{\alpha}(k)$
varies continuously, and thus the equations can describe a pulse of
light with arbitrary polarization incident on only a finite region
of the structure shown in Fig.~\ref{fig:graphene}. Here we consider
simple cases with incident plane waves at discrete $k_{j}$, which
is performed by the transformation 
\begin{eqnarray}
{E}_{\text{cl};-}^{\alpha}(k)\to(2\pi)^{3}\sum_{j}\delta(k-k_{j}){E}_{\text{cl};-}^{\alpha}(k_{j})\,.
\end{eqnarray}
All new $k_{j}$ generated by the nonlinearity are then also discrete.

In the following two subsections, we explicitly give the perturbation
formulas for SHG, THG, and PFC.

\subsection{SHG and THG\label{sec:hg}}

In this section we derive the formulas for the SHG and THG signals,
induced by a single $p-$polarized incident laser beam with amplitude
${E}_{\text{cl};-}^{p}(k)$, as shown in Fig.~\ref{fig:graphene}. We list all perturbative quantities with
respect to the incident fields. The perturbative effective fields
at the graphene layer are given as 
\begin{align*}
\bm{E}_{\text{eff};\parallel}^{(1)}(k) & ={\cal P}(k)E_{\text{cl};-}^{p}(k)\hat{\bm{\kappa}}\,,\\
\bm{E}_{\text{eff};\parallel}^{(2)}(k) & =-{\cal P}(k){\cal J}_{\text{nl};\kappa}^{(2)}(k)\hat{\bm{\kappa}}\,,
\end{align*}
with 
\begin{equation}
{\cal P}(k)=[1-R_{cs}^{p}(k)]\frac{w_{0}(k)}{\widetilde{\omega}}\,.\label{eq:Pkdef}
\end{equation}
The perturbative current responses for SHG and THG are \begin{subequations}
\begin{align}
\vec{{\cal J}}_{\text{nl};\parallel}^{(2)}(2k) & =\eta_{2}(k,k)\hat{\bm{\kappa}}[E_{\text{eff};\kappa}^{(1)}(k)]^{2}\,,\label{eq:jnlshggr}\\
\vec{{\cal J}}_{\text{nl};\parallel}^{(3)}(3k) & =\eta_{3}(\omega,\omega,\omega)\hat{\bm{\kappa}}[E_{\text{eff};\kappa}^{(1)}(k)]^{3}\nonumber \\
 & +2\eta_{2}(k,2k)\hat{\bm{\kappa}}E_{\text{eff};\kappa}^{(1)}(k)E_{\text{eff};\kappa}^{(2)}(2k)\,.\label{eq:jnlthggr}
\end{align}
\end{subequations}
Substituting them into Eq.~(\ref{eq:outE}), the radiation fields for SHG and THG
only include the $p$-polarized components as \begin{subequations}
\begin{align}
E_{\text{cl};+}^{p}(2k) & =\eta_{\text{SHG}}(k)\beta_{\text{SHG}}(k)[E_{\text{cl};-}^{p}(k)]^{2}\,,\\
E_{\text{cl};+}^{p}(3k) & =\eta_{\text{THG}}(k)\beta_{\text{THG}}(k)[E_{\text{cl};-}^{p}(k)]^{3}\,,
\end{align}
\end{subequations} with $\eta_{\text{SHG}}(k)=\eta_{2}(k,k)$ and
\begin{equation}
\eta_{\text{THG}}(k)=\eta_{3}(\omega,\omega,\omega)-2\eta_{2}(k,2k){\cal P}(2k)\eta_{2}(k,k)\,.\label{eq:etathg}
\end{equation}
Here all quantities indicated by $\beta$ are dimensionless structure
factors, which include the effects from the lower layers of the structure.
They are given by
\begin{subequations} 
\begin{align}
\beta_{\text{SHG}}(k) & =\left[1-R_{cs}^{p}(2k)\right][{\cal P}(k)]^{2}\,,\\
\beta_{\text{THG}}(k) & =\left[1-R_{cs}^{p}(3k)\right][{\cal P}(k)]^{3}\,.
\end{align}
\end{subequations}

\begin{figure}[!htpb]
\centering \includegraphics[width=7cm]{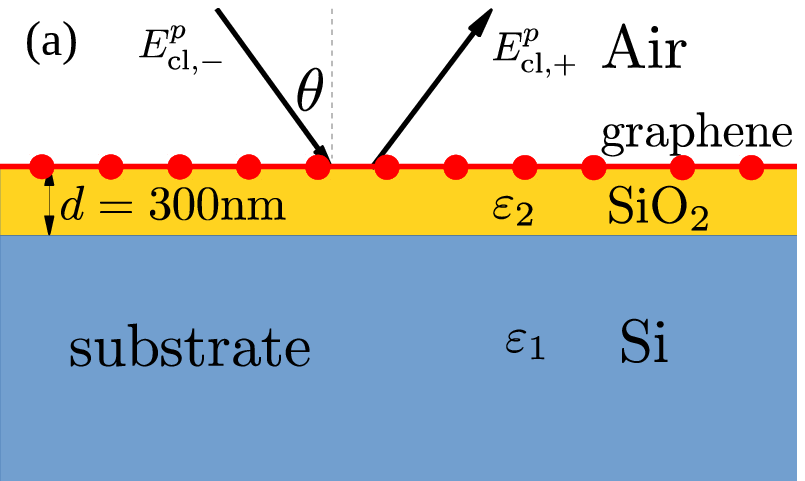}\\
 \includegraphics[width=7cm]{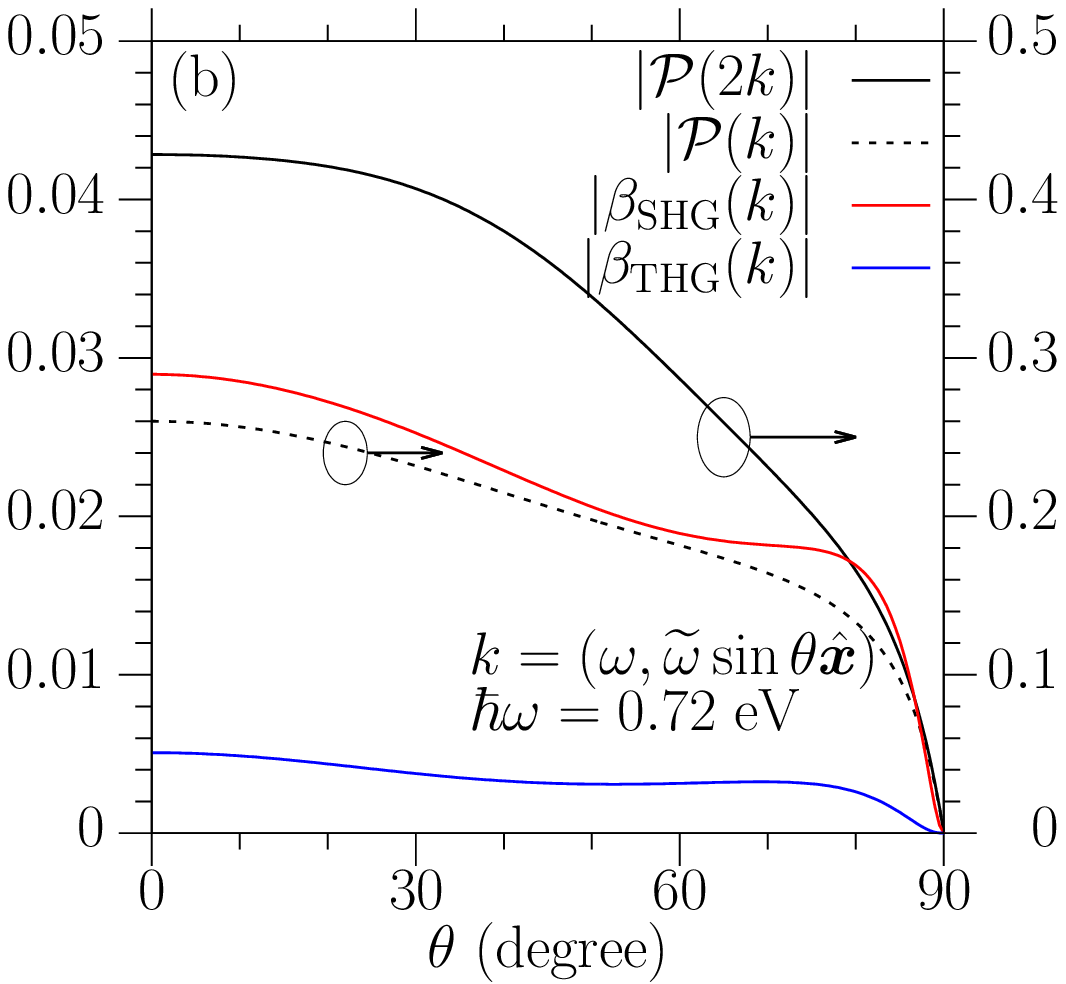}
\caption{(color online) (a) Illustration of the structure analyzed, (b) the
angle dependence of $|\beta_{\text{SHG}}(k)|$, $|\beta_{\text{THG}}(k)|$,
$|{\cal P}(2k)|$, and $|{\cal P}(k)|$. The $y$-axis of $|{\cal P}|$s
is at the right, as indicated. }
\label{fig:thg_zhao} 
\end{figure}

As an example, we consider these structure related quantities for
a THG experimental setup as used by Kumar \textit{et al.} \cite{Phys.Rev.B_87_121406_2013_Kumar}.
The layer structure is shown in Fig.~\ref{fig:thg_zhao}(a); it
is composed of air cladding/graphene/300~nm thick SiO$_{2}$/silicon
substrate. The incident light is at $\hbar\omega=0.72$~eV, with
an incident angle $\theta$ or in-plane wavevector $\bm{\kappa}=\widetilde{\omega}\sin\theta\hat{\bm{x}}$.
The dielectric constants we used are listed in Table~\ref{tab:thgpara},
and the linear conductivity of graphene is taken as $\sigma_{0}$.

\begin{table}[!htpb]
\centering %
\begin{tabular}[t]{ccccc}
\hline 
 & $\varepsilon(\omega)$  & $\varepsilon(2\omega)$ & $\varepsilon(3\omega)$ & \tabularnewline
\hline 
SiO$_{2}$  & 2.08  & 2.11  & 2.13  & \tabularnewline
Si  & 12.01  & $13.27+0.022i$  & $16.03+0.189i$  & \tabularnewline
\hline 
\hline 
 & $R_{cs}^{p}(k_{\perp})$  & $R_{cs}^{p}(2k_{\perp})$  & $R_{cs}(3k_{\perp})$ & \tabularnewline
\hline 
 & $0.74+0.001i$  & $0.57-0.007i$ & $0.71+0.006i$ & \tabularnewline
\hline 
\hline 
\multicolumn{4}{c}{graphene} & \tabularnewline
\hline 
\multicolumn{2}{c}{ $\eta_{2}(k,k)/\sin\theta$} & \multicolumn{2}{c}{$(1.1-0.6i)\times10^{-15}$ m/V} & \tabularnewline
\multicolumn{2}{c}{$\eta_{2}(k,2k)/\sin\theta$ } & \multicolumn{2}{c}{$(1.2-0.53)\times10^{-16}$ m/V} & \tabularnewline
\hline 
\multicolumn{2}{c}{ $\eta_{3}(\omega,\omega,\omega)$} & \multicolumn{2}{c}{$(6.4+1.6i)\times10^{-22}$ m$^{2}$/V$^{2}$} & \tabularnewline
\hline 
\end{tabular}\caption{The values for the dielectric constants of SiO$_{2}$ and Si (taken
from website `http://refractiveindex.info/''), the total reflection
coefficient at normal incidence, and the optical conductivities of
graphene. The nonlinear conductivities of graphene are taken from
earlier
work~{[}\protect\onlinecite{Phys.Rev.B_91_235320_2015_Cheng,*Phys.Rev.B_93_39904_2016_Cheng},
\protect\onlinecite{arXiv:1609.06413}{]}, with parameters $\mu=0.2$~eV,
$\Gamma_{e}=\Gamma_{i}=33$~meV, and $T=300$~K. The linear conductivity
of graphene is approximated as $\sigma_{0}$. }
\label{tab:thgpara} 
\end{table}

In Fig.~\ref{fig:thg_zhao} (b), we plot the $\theta$ dependence
of $|\beta_{\text{SHG}}(k)|$, $|\beta_{\text{THG}}(k)|$, and $|{\cal P}(2k)|$.
As the incident angle is increased, they decrease. At normal incidence,
the values of $|\beta_{\text{SHG}}|$ and $|\beta_{\text{THG}}|$
are of the order of $10^{-2}$ and $10^{-3}$, respectively. Both
of them are much smaller than $1$, which means the layer structure
greatly reduces the harmonic radiation signal. This is because the
value of $|{\cal P}(k)|$ is less than 0.3, and as it is raised to
a power of 2 and 3 the resulting values are much smaller. At normal
incidence $k=k_\perp=(\bm 0,\omega)$,  the structure factors can be
written as $\beta_{\text{SHG}}(k_{\perp})={\cal  P}(2k_{\perp})[{\cal P}(k_{\perp})]^{2}\approx2.9\times10^{-2}$ 
and $\beta_{\text{THG}}(k_{\perp})={\cal P}(3k_{\perp})[{\cal P}(k_{\perp})]^{3}\approx5\times10^{-3}$.
Compared to a suspended graphene, where both of them are around unity, 
the layers underneath reduce the harmonic radiation
significantly. The dependence of these structure factors on the substrate reflection
coefficient is clear.

Equation~(\ref{eq:etathg}) includes the contribution to the THG coefficients
  from the cascaded processes, with the values of $|{\cal P}(2k)|$ smaller than 0.5. By checking graphene's conductivities listed in Table~\ref{tab:thgpara},
we find the contribution from the cascaded second order processes
to the THG process are about 9 orders of magnitude smaller than the
direct third order response. This is not surprising, because the second
order response in graphene is induced by theoretically forbidden optical
processes and is a weak effect, which gives very small $\eta_{2}(k,k)$
and $\eta_{2}(k,2k)$. And ${\cal P}(2k)$ does not undergo any resonance
for the structure and parameters used in the experiment.

Following a similar strategy, in the next section we look at
parametric frequency conversion. 

\subsection{Parametric frequency conversion\label{sec:pfc}}

\begin{figure}[!htp]
\centering \includegraphics[width=7cm]{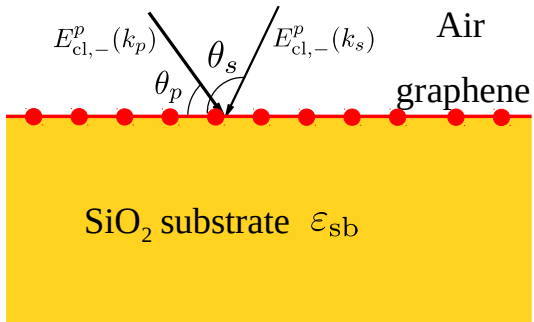}
\caption{(color online) Illustration of the structure in the experiment of
Constant \textit{et al.}\cite{Nat.Phys._12_124_2015_Constant1}. }
\label{fig:hendry} 
\end{figure}
The PFC process occurs when the structure is excited by two laser beams: one is a pump beam at $k_{p}$, the
other is a signal beam at $k_{s}$. We are interested in the output
light at a difference frequency $k_{d}=k_{p}-k_{s}$, which is generated
by the second order nonlinearity, and in the idler light at
$k_{i}=2k_{p}-k_{s}$, 
which is generated by the third order nonlinearity. In the experiment
of Constant \textit{et al.} \cite{Nat.Phys._12_124_2015_Constant1},
due to a carefully designed structure and appropriate incident angles
the light at $k_{d}$ can be evanescent and on resonance with the
plasmon modes of the whole structure. As shown in Fig.~\ref{fig:hendry},
the structure is composed of graphene covered SiO$_{2}$. For simplicity,
we consider that the $p$-polarized incident beams are in the same
incident plane, 
with in-plane wave vectors $\bm{\kappa}_{p}=\kappa_{p}\hat{\bm{\kappa}}_{p}$
and $\bm{\kappa}_{s}=\kappa_{s}(-\hat{\bm{\kappa}}_{p})$, and with
frequencies $\omega_{p}>\omega_{s}$.

In a calculation similar to that of the previous section, we find
the output fields at $k_{d}$ and $k_{i}$ are
\begin{subequations}
\begin{align}
E_{\text{cl};+}^{p}(k_{d}) & =-2\eta_{\text{DFG}}(k_{p},k_{s})\beta_{\text{DFG}}(k_{p},k_{s})\nonumber \\
 & \times E_{\text{cl};-}^{p}(k_{p})[E_{\text{cl};-}^{p}(k_{s})]^{\ast}\,,\\
E_{\text{cl};+}^{p}(k_{i}) & =-3\eta_{\text{PFC}}(k_{p},k_{s})\beta_{\text{PFC}}(k_{p},k_{s})\nonumber \\
 & \times[E_{\text{cl};-}^{p}(k_{p})]^{2}[E_{\text{cl};-}^{p}(k_{s})]^{\ast}\,,
\end{align}
\end{subequations}
The prefactors $2$ and $3$ come from the permutation
of the incident field components, and the minus signs appear due
  to the opposite incident directions of these two beams. The terms related to the conductivities
are $\eta_{\text{DFG}}(k_{p},k_{s})=\eta_{2}(k_{p},-k_{s})$ and $\eta_{\text{PFC}}(k_{p},k_{s})=\eta_{3}(\omega_{p},\omega_{p},-\omega_{s})+\eta_{3c}(k_{p},k_{s})$.
The term $\eta_{3c}$ includes the contribution from the cascaded
second order processes to the third order response, and is given as
\begin{align}
\eta_{3c}(k_{p},k_{s}) & =-\frac{4}{3}\eta_{2}(k_{p},k_{d}){\cal P}(k_{d})\eta_{2}(k_{p},-k_{s})\,.\label{eq:eta3c}
\end{align}
Here only the cascaded process involving the light at $k_{d}$ is taken
into account. The dimensionless structure factors are \begin{subequations}
\begin{align}
\beta_{\text{DFG}}(k_{p},k_{s}) & =[1-R_{cs}^{p}(k_{d})]{\cal P}(k_{p})[{\cal P}(k_{s})]^{\ast}\,,\\
\beta_{\text{PFC}}(k_{p},k_{s}) & =[1-R_{cs}^{p}(k_{i})][{\cal P}(k_{p})]^{2}[{\cal P}(k_{s})]^{\ast}\,.
\end{align}
\end{subequations}

\begin{widetext}
\begin{figure*}[!htpb]
\centering \includegraphics[width=17cm]{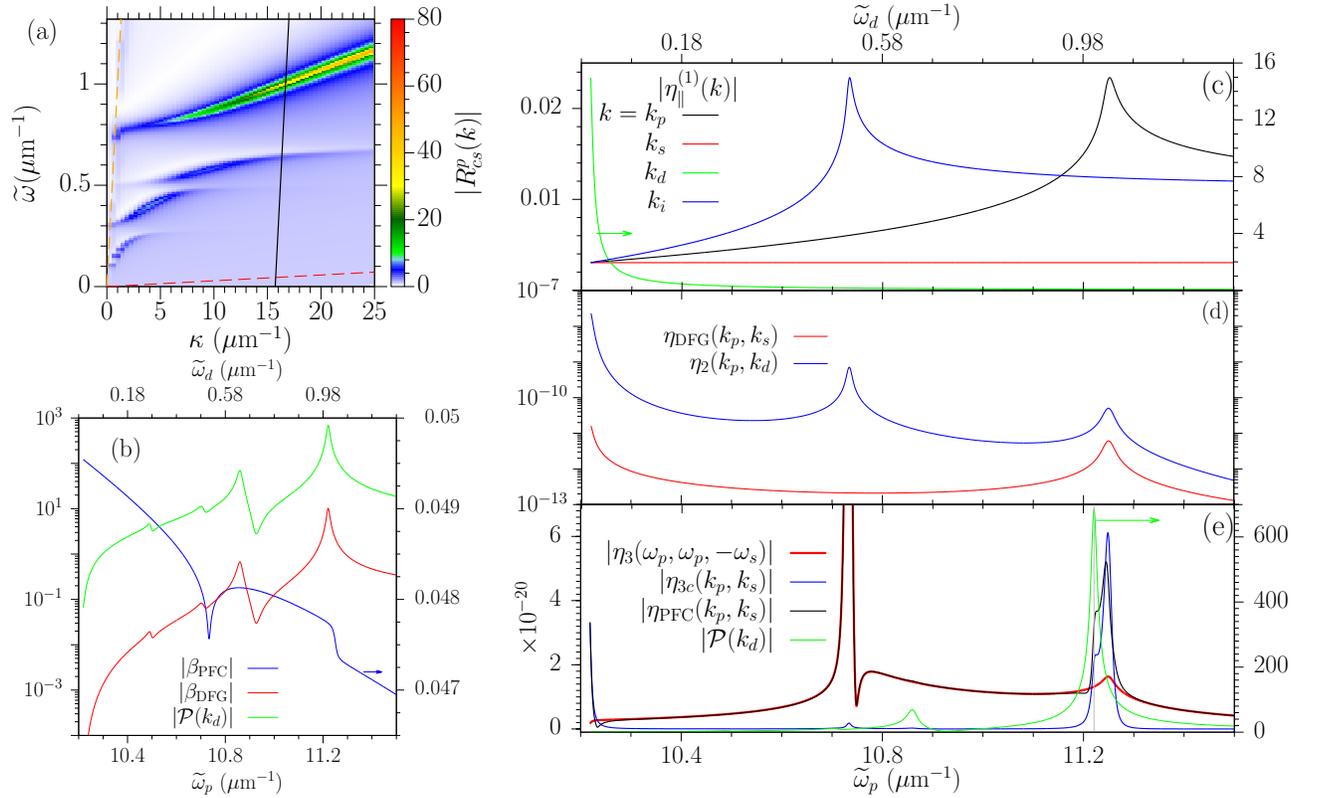}
\caption{(color online) (a) $k$-dependence of the reflection coefficients.
The orange dashed line is the light line, the red dashed line is the
graphene dispersion, and the black line is the trajectory of $k_{d}$
when $\lambda_{p}$ varies from 547~nm to 615~nm for $\lambda_{s}=615$~nm,
$\theta_{s}=125^{\circ}$, and $\theta_{p}=15^{\circ}$. (b) The structure
factors $\beta_{\text{PFC}}(k_{p},k_{s})$, $\beta_{\text{DFG}}(k_{p},k_{s})$,
and ${\cal P}(k_{d})$. (c) $\eta_{\parallel}^{(1)}(k)$ for $k=k_{p}$,
$k_{s}$, $k_{d}$, and $k_{i}$. (d) $\eta_{\text{DFG}}(k_{p},k_{s})$
and $\eta_{2}(k_{p},k_{d})$. (e) $\eta_{\text{PFC}}(k_{p},k_{s})$,
$\eta_{3c}(k_{p},k_{s})$, $\eta_{3}(\omega_{p},\omega_{p},-\omega_{s})$,
and ${\cal P}(k_{d})$. All curves indicated by horizontal arrows use
the right $y$-axis. In the calculation, the parameters for graphene are taken
as $\mu=1.11$~meV, $\Gamma_{i}=\Gamma_{e}=1$~meV, and $T=0$~K.}
\label{fig:fwm} 
\end{figure*}
\end{widetext}

Besides involving a product of two second order conductivities, the
cascaded process coefficient $\eta_{3c}(k_{p},k_{s})$ in Eq.~(\ref{eq:eta3c})
is also proportional to the term ${\cal P}(k_{d})$. This
illustrates how the structure underneath the graphene can be used
to tune the cascaded process: In the experiment by Constant \textit{et
al.} \cite{Nat.Phys._12_124_2015_Constant1}, this term is maximized
by matching $k_{d}$ with the surface plasmon polariton (SPP) resonance 
of the entire structure. In an ideal case, when all losses are ignored,
$R_{cs}^{p}(k_{d})$ diverges at the resonance; with losses, its value
becomes finite, but still very large. As an illustration, this can be
clearly seen in Fig.~\ref{fig:fwm}(a). For the parameters of
graphene we choose the chemical potential as $\mu=1.11$~eV, the
relaxation parameters $\Gamma_{i}=\Gamma_{e}=1$~meV, and assume
zero temperature. The very high chemical potential, which can be realized
by a gate voltage \cite{Nat.Photon.___2018_Jiang}, is chosen to satisfy resonant conditions
in calculating the second order and third order conductivities. The
calculation is made for signal light at wavelength $615$~nm ($\widetilde{\omega}_{s}=10.22~\mu$m$^{-1}$)
with an incident angle $\theta_{s}=125^{\circ}$, and pump light
at wavelength $\lambda_{p}$ varying from $547$~nm to $615$~nm
($\widetilde{\omega}_{p}$ varying from $11.49~\mu$m$^{-1}$ to $10.22~\mu$m$^{-1}$)
with an incident angle $\theta_{p}=15^{\circ}$. As the pump light
wavelength varies, the in-plane wave vector and frequency of the light
at $k_{d}=k_{p}-k_{s}$ vary along the black line in Fig.~\ref{fig:fwm}(a).
The linear conductivities for graphene are shown in Fig.~\ref{fig:fwm}(c).
The maximal value of $|R_{cs}^{p}(k_{d})|$ along the black line in
Fig.~\ref{fig:fwm}(a) is about $41$, and occurs for $\widetilde{\omega}_{p}=11.22~\mu$m$^{-1}$where
the SPP is excited.

We first check the structure factor for PFC, 
which is shown in Fig.~\ref{fig:fwm}(b). As $\lambda_{p}$ varies,
$|\beta_{\text{PFC}}|$ is at the order of 0.05 and changes only slightly.
This small value results mostly because of the oblique incidence of
the light, as captured by the term $w_{0}(k)/\widetilde{\omega}$
in the expression of Eq.~(\ref{eq:Pkdef}) for ${\cal P}(k)$. For the difference
frequency generation, the structure factor $|\beta_{\text{DFG}}|$
can be as large as $10$ at the SPP resonance at $\widetilde{\omega}_{p}=11.22\mu$m$^{-1}$.
Obviously, the light at $k_{d}$ is evanescent in air; however, it
is could be detected by the use of near field techniques.

Interestingly, the cascaded second order processes can exceed the
direct third order process at certain frequencies, mainly because in
this case we have very large second order conductivities as well as a
SPP resonance, as shown in Fig.~\ref{fig:fwm} (d) and \ref{fig:fwm} (e). There are in total three
 possible resonances for the conductivities with
 varying
$\widetilde{\omega}_{p}$, the first is at $\widetilde{\omega}_{A}=[\widetilde{\omega}_{s}+2\mu/(\hbar c)]/2=10.73\mu$m$^{-1}$,
corresponding to the condition $2\hbar\omega_{A}-\hbar\omega_{s}=2\mu$;
the second is at $\widetilde{\omega}_{B}=\widetilde{\omega}_{s}$;
and the third is at $\widetilde{\omega}_{C}=2\mu/(\hbar c)=11.25\mu$m$^{-1}$,
corresponding to the condition $\hbar\omega_{C}=2\mu$. Our calculations
show that $\eta_{2}(k_{p},-k_{s})$ shows peaks at $\widetilde{\omega}_{B}$
and $\widetilde{\omega}_{C}$; $\eta_{2}(k_{p},k_{d})$ shows peaks
at $\widetilde{\omega}_{A}$, $\widetilde{\omega}_{B}$, and $\widetilde{\omega}_{C}$;
and $\eta_{3}(\omega_{p},\omega_{p},-\omega_{s})$ shows peaks at
$\widetilde{\omega}_{A}$ and $\widetilde{\omega}_{c}$. For $\widetilde{\omega}_{p}$
around $\widetilde{\omega}_{B}$ and $\widetilde{\omega}_{C}$, the
cascaded second order processes dominate because of the very large
second order conductivities. But there is another peak for $\widetilde{\omega}_{p}\sim11.22~\mu$m,
located at the same peak position of ${\cal P}(k_{d})$, which is
induced by the SPP resonance.

All these calculations are for ideal parameters. For higher temperature,
larger relaxation parameters, and smaller chemical potential, the
second order conductivities can be reduced greatly, as well the value
of $\eta_{3c}(k_{p},k_{s})$. Because the graphene layer is put in
an asymmetric environment, an additional contribution to the second
order nonlinearity can arise from the interface effect. However, it
contributes only to the out-of-plane tensor components, which are ignored
in the usual model for graphene.

\section{Conclusion\label{sec:conclude}}

We have derived the basic equations for analyzing the linear and nonlinear
optics of graphene and other 2D materials in layered structures. For
linear optics, we obtained the Fresnel coefficients of a suspended
2D layer. Complementing existing work in the literature, these coefficients
take into account the optical response along the out-of-plane direction
of the 2D layers, which is usually ignored, so they can be used to
extract the out-of-plane linear conductivity components from experimental
data. In the nonlinear regime, we derived an equation for the total
electric field at the 2D material including the feedback radiation of its
own sheet current density. After combining this with current response
theory, which is widely adopted for calculating the conductivities,
a set of consistent equations for the total field and the current
can be obtained; after solving them either analytically or numerically, the
nonlinear radiation can be found. We also derived the perturbative
expressions at weak incident light beams. We discussed how the structures
themselves affect the output field intensity. For some experimental
scenarios the structures can reduce the radiation by several orders
of magnitude, compared to that for a suspended sample. Our results
can serve as a starting point for further nonlinear optical investigations
of graphene or 2D materials in layered structures.

\acknowledgements This work has been supported by CAS
QYZDB-SSW-SYS038, NSFC Grant No. 11774340 and 61705227. N.V. acknowledges the ERC-FP7/2007-2013 grant 336940, the
Research Fundation Flanders (FWO) under Grants G.A002.13N and G0F6218N
(EOS-Convention 30467715), VUB-Methusalem, and VUB-OZR. J.E.S. is
supported by the Natural Sciences and Engineering Research Council of
Canada.

\appendix

\section{Notations \label{app:notation}}

In a homogeneous medium with dielectric constant $\varepsilon(\omega)$,
the electric fields can be written as 
\begin{eqnarray}
\bm{E}(z;\bm{R},t) & = & \int_{0}^{\infty}\frac{d\omega}{2\pi}\bm{E}(z;\bm{R},\omega)e^{-i\omega t}+c.c.\,,\\
\bm{E}(z;\bm{R},\omega) & = & \int\frac{d\bm{\kappa}}{(2\pi)^{2}}\bm{E}(z;\bm{\kappa},\omega)e^{i\bm{\kappa}\cdot\bm{R}}\,,
\end{eqnarray}
with $\bm{R}=x\hat{\bm{x}}+y\hat{\bm{y}}$. Throughout this work,
we use a compact notation $k=(\bm{\kappa},\omega)$, and so $\bm{E}(z;\bm{\kappa},\omega)=\bm{E}(z;k)$.
The fields can be decomposed into upward and downward propagating
(or evanescent) fields, 
\begin{equation}
{\bm{E}}(z;k)=\sum_{\lambda=\pm}{\bm{E}}_{\lambda}(k)e^{i\lambda w(k)z}\,,
\end{equation}
with $s$- and $p$-polarized components 
\begin{equation}
{\bm{E}}_{\lambda}(k)=\sum_{\alpha=s,p}{E}_{\lambda}^{\alpha}(k)\hat{\bm{e}}_{\lambda}^{\alpha}(k)\,.\label{eq:eprop}
\end{equation}
The polarization vectors, $\hat{\bm{e}}_{\lambda}^{\alpha}(k)$ with
$\alpha=s,p$ for the field polarization and $\lambda=+,-$ for the
propagation direction, are defined by 
\begin{eqnarray}
\hat{\bm{e}}_{\pm}^{s}(k) & \equiv & \hat{\bm{s}}(k)=\hat{\bm{\kappa}}\times\hat{\bm{z}}\,,\\
\hat{\bm{e}}_{\pm}^{p}(k) & \equiv & \hat{\bm{p}}_{\pm}(k)=\frac{\kappa\hat{\bm{z}}\mp w(k)\hat{\bm{\kappa}}}{\sqrt{\varepsilon(\omega)}\widetilde{\omega}}\,.
\end{eqnarray}
where $w(k)=\sqrt{\varepsilon(\omega)\widetilde{\omega}^{2}-\kappa^{2}}$
with $\widetilde{\omega}=\omega/c$, is taken such that $\text{Im}[w]\geq0$,
with $\text{Re}[w]>0$ if $\text{Im}[w]=0$. The polarization vectors
are in general complex quantities; for propagating fields in lossless
materials, their directions are shown in Fig.~\ref{fig:2Doptic}.
Although these vectors satisfy $\hat{\bm{s}}(k)\cdot\hat{\bm{s}}(k)=1$
and $\hat{\bm{p}}_{\pm}(k)\cdot\hat{\bm{p}}_{\pm}(k)=1$, the ``length''
$\sqrt{\hat{\bm{p}}_{\pm}(k)\cdot[\hat{\bm{p}}_{\pm}(k)]^{\ast}}$
can be far larger than unity for an evanescent field.

To identify different regions or layers, we will use an extra subscript
$j$ in $E_{j;\lambda}^{\alpha}(z;k)$, $w_{j}(k)$, $\hat{\bm{e}}_{j;\lambda}^{\alpha}(k)$,
and so on.

\section{Green function method for a sheet current density\label{app:gfunc}}

For a sheet current density $\bm{J}(z;k)$ embedded in a homogeneous
medium with dielectric constant $\varepsilon(\omega)$, the radiation
field can be calculated using a Green function strategy \cite{J.Opt.Soc.Am.B_4_481_1987_Sipe}
as 
\begin{equation}
\bm{E}_{\text{ind}}(z;k)=\int G(z-z^{\prime};k)\cdot\frac{\bm{J}(z^{\prime};k)}{2c\epsilon_{0}}dz^{\prime}\,,\label{eq:eind}
\end{equation}
with 
\begin{eqnarray}
G(z;k) & = & -\frac{\tilde{\omega}}{w(k)}[\hat{\bm{s}}(k)\hat{\bm{s}}(k)+\hat{\bm{p}}_{+}(k)\hat{\bm{p}}_{+}(k)]\theta(z)e^{iw(k)z}\nonumber \\
 & - & \frac{\tilde{\omega}}{w(k)}[\hat{\bm{s}}(k)\hat{\bm{s}}(k)+\hat{\bm{p}}_{-}(k)\hat{\bm{p}}_{-}(k)]\theta(-z)e^{-iw(k)z}\nonumber \\
 & - & 2i\frac{\hat{\bm{z}}\hat{\bm{z}}}{\widetilde{\omega}\varepsilon(\omega)}\delta(z)\,,\label{eq:mwg}
\end{eqnarray}
In the region $z>0$ of the setup in Fig.~\ref{fig:2Doptic}, the
induced field $\bm{E}_{\text{ind}}(z>0;k)$ only propagate upwards, and
it can be written as 
\begin{equation}
  \bm{E}_{\text{ind}}(z>0;k) = \sum_{\alpha=s,p} {\cal F}_{+}^\alpha(k)
  e^{iw(k)z}\hat{\bm e}_{+}^\alpha(k)
\end{equation}
with ${\cal F}_+^\alpha$ given in Eq.~(\ref{eq:calF2D}). Then the upward 
fields $\bm E_{1;+}(k)$ include two contributions: one is from the
incident fields $\bm E_{2;+}(k)$ that is propagated from the region
$z<0$, as if the 
sheet current density were absent; the other is the induced field
${\cal F}_+^\alpha(k)$. Then we get Eq.~(\ref{eq:eu}). Similarly, we
can get Eq.~(\ref{eq:ed}).

\section{Incident field $\bm{E}_{\text{eff}}(k)$ at the 2D layer and
  the radiation coefficients ${\cal Q}(k)$ \label{app:factor}}

For a 2D layer inside a multilayered structure, it is convenient to
write the incident field $\bm{E}_{\text{2D}}$ at the 2D layer in
a coordinate formed by $\{\hat{\bm{s}},\hat{\bm{\kappa}},\hat{\bm{z}}\}$
as 
\begin{equation}
\bm{E}_{\text{2D}}=E_{\text{2D};s}\hat{\bm{s}}+E_{\text{2D};\kappa}\hat{\bm{\kappa}}+E_{\text{2D};z}\hat{\bm{z}}\,.
\end{equation}
Without causing any confusion, the $k$-dependence of each quantity
is implicitly shown. From Eqs.~(\ref{eq:eu}), (\ref{eq:ed}), (\ref{eq:totE2D}),
(\ref{eq:TMu}), and (\ref{eq:TMd}), the effective field
$\bm{E}_{\text{eff}}$ at the 2D layer can be written into Eq.~(\ref{eq:radE1})
with 
\begin{widetext}
  {\allowdisplaybreaks
\begin{subequations} 
\begin{align}
E_{\text{2D};s} & =\frac{[\overline{R}_{cs}^{s}-R_{cu}^{s}][1+R_{uc}^{s}]+T_{cu}^{s}T_{uc}^{s}}{T_{cu}^{s}}E_{\text{cl};-}^{s}+\frac{[1+R_{uc}^{s}]\overline{T}_{cs}^{s}}{T_{cu}^{s}}E_{\text{sb};+}^{s}\,,\\
E_{\text{2D};\kappa} & =\left\{ \frac{-[\overline{R}_{cs}^{p}-R_{cu}^{p}][1-R_{uc}^{p}]+T_{cu}^{p}T_{uc}^{p}}{T_{cu}^{p}}E_{\text{cl};-}^{p}+\frac{[-1+R_{uc}^{p}]\overline{T}_{cs}^{p}}{T_{cu}^{p}}E_{\text{sb};+}^{p}\right\} \frac{w_{0}}{\widetilde{\omega}}\,,\\
E_{\text{2D};z} & =\left\{ \frac{[\overline{R}_{cs}^{p}-R_{cu}^{p}][1+R_{uc}^{p}]+T_{cu}^{p}T_{uc}^{p}}{T_{cu}^{p}}E_{\text{cl};-}^{p}+\frac{[1+R_{uc}^{p}]\overline{T}_{cs}^{p}}{T_{cu}^{p}}E_{\text{sb};+}^{p}\right\} \frac{\kappa}{\widetilde{\omega}}\,.
\end{align}
\end{subequations} }
\end{widetext}
Here $\overline{T}^{\alpha}$ and $\overline{R}^{\alpha}$ are the
Fresnel coefficient for the structure without the presence of 2D material,
which are given from {\allowdisplaybreaks \begin{subequations}
\begin{align}
{\cal M}_{cu}^{\alpha}{\cal M}_{ls}^{\alpha}=\frac{1}{\overline{T}_{sc}^{\alpha}}\begin{pmatrix}\overline{T}_{cs}^{\alpha}\overline{T}_{sc}^{\alpha}-\overline{R}_{cs}^{\alpha}\overline{R}_{sc}^{\alpha} & \overline{R}_{cs}^{\alpha}\\
-\overline{R}_{sc}^{\alpha} & 1
\end{pmatrix}\,;
\end{align}
while $R$ and $T$ are the Fresnel coefficients for the substrate
multilayers, which are given as 
\begin{align}
{\cal M}_{cu}^{\alpha} & =\frac{1}{{T}_{uc}^{\alpha}}\begin{pmatrix}{T}_{cu}^{\alpha}{T}_{uc}^{\alpha}-{R}_{cu}^{\alpha}{R}_{uc}^{\alpha} & {R}_{cu}^{\alpha}\\
-{R}_{uc}^{\alpha} & 1
\end{pmatrix}\,.
\end{align}
\end{subequations} Along the directions $\{\hat{\bm{s}},\hat{\bm{\kappa}},\hat{\bm{z}}\}$,
  the radiation coefficient is a tensor matrix with a form
  ${\cal Q}={\cal Q}_{ss}\hat{\bm s}\hat{\bm s}+{\cal
    Q}_{\kappa\kappa}\hat{\bm\kappa}\hat{\bm\kappa}+{\cal
    Q}_{zz}\hat{\bm z}\hat{\bm z}+{\cal Q}_{\kappa
    z}\hat{\bm\kappa}\hat{\bm z}+{\cal Q}_{z\kappa}\hat{\bm z}\hat{\bm\kappa}$
  with elements 
\begin{subequations} 
\begin{align}
{\cal Q}_{ss} & =-\frac{\widetilde{\omega}}{w_{0}}\frac{[\overline{R}_{cs}^{s}-R_{cu}^{s}][1+R_{uc}^{s}]+T_{cu}^{s}T_{uc}^{s}}{T_{cu}^{s}}\frac{1+R_{uc}^{s}}{T_{uc}^{s}}\,,\\
{\cal Q}_{\kappa\kappa} & =-\frac{w_{0}}{\widetilde{\omega}}\frac{-[\overline{R}_{cs}^{p}-R_{cu}^{p}][1-R_{uc}^{p}]+T_{cu}^{p}T_{uc}^{p}}{T_{cu}^{p}}\frac{1-R_{uc}^{p}}{T_{uc}^{p}}\,,\\
{\cal Q}_{zz} & =-\frac{\kappa^{2}}{w_{0}\widetilde{\omega}}\frac{[\overline{R}_{cs}^{p}-R_{cu}^{p}][1+R_{uc}^{p}]+T_{cu}^{p}T_{uc}^{p}}{T_{cu}^{p}}\frac{1+R_{uc}^{p}}{T_{uc}^{p}}\,,\\
{\cal Q}_{\kappa z} & =-{\cal Q}_{z\kappa}=-\frac{\kappa}{\widetilde{\omega}}\frac{[R_{cu}^{p}-\overline{R}_{cs}^{p}]\{1-[R_{uc}^{p}]^{2}\}+R_{uc}^{p}T_{cu}^{p}T_{uc}^{p}}{T_{cu}^{p}T_{uc}^{p}}\,.
\end{align}
\end{subequations} 

We give these quantities for two special cases:\\
 (1) For a suspended system, all reflection coefficients associated
with additional material are zero and all transmission coefficients
are $1$. Therefore we can get the effective incident field at the
2D layer is \begin{subequations} 
\begin{align}
\bm{E}_{\text{2D}}^{0} & =[E_{1;-}^{s}+E_{2;+}^{s}]\hat{\bm{s}}
 +\frac{w_{0}}{\widetilde{\omega}}\left[E_{1;-}^{p}-E_{2;+}^{p}\right]\hat{\bm{\kappa}}\nonumber \\
 & +\frac{\kappa}{\widetilde{\omega}}\left[E_{1;-}^{p}+E_{2;+}^{p}\right]\hat{\bm{z}}\,,
\end{align}
and the radiation coefficient matrix becomes diagonal,
\begin{equation}
  {\cal Q}^0=-\frac{\widetilde{\omega}}{w_{0}} \hat{\bm s}\hat{\bm s}
  -\frac{w_{0}}{\widetilde{\omega}}\hat{\bm\kappa}\hat{\bm\kappa}
  -\frac{\kappa^{2}}{w_{0}\widetilde{\omega}} \hat{\bm z}\hat{\bm z}\,.
\end{equation}
\end{subequations}
 (2) When the 2D layer is put directly on top of a substrate, we have
$R_{cu}^{\alpha}=R_{uc}^{\alpha}=0$, and $T_{cu}^{\alpha}=T_{uc}^{\alpha}=1$.
The effective field is \begin{subequations} 
\begin{align}
E_{\text{2D};s} & =[1+R_{ls}^{s}]E_{\text{cl};-}^{s}+T_{ls}^{s}E_{\text{sb};+}^{s}\,,\label{eq:e2dsgh}\\
E_{\text{2D};\kappa} & =\left\{ [1-R_{ls}^{p}]E_{\text{cl};-}^{p}-T_{ls}^{p}E_{\text{sb};+}^{s}\right\} \frac{w_{0}}{\widetilde{\omega}}\,,\label{eq:e2dkgh}\\
E_{\text{2D};z} & =\left\{ [1+R_{ls}^{p}]E_{\text{cl};-}^{p}+T_{ls}^{p}E_{\text{sb};+}^{s}\right\} \frac{\kappa}{\widetilde{\omega}}\,,
\end{align}
\end{subequations} with $\overline{R}_{cs}^{\alpha}=R_{ls}^{\alpha}$
and $\overline{T}_{cs}^{\alpha}=T_{ls}^{\alpha}$ being the Fresnel
coefficient associated with the lower multilayered structure. The
${\cal Q}$ is then given as \begin{subequations} 
\begin{align}
{\cal Q}_{ss} & =-\frac{\widetilde{\omega}}{w_{0}}\left[1+R_{ls}^{s}\right]\,,\label{eq:qssgh}\\
{\cal Q}_{\kappa\kappa} & =-\frac{w_{0}}{\widetilde{\omega}}\left[1-R_{ls}^{p}\right]\,,\label{eq:qkkgh}\\
{\cal Q}_{zz} & =-\frac{\kappa^{2}}{w_{0}\widetilde{\omega}}\left[R_{ls}^{p}+1\right]\,,\\
{\cal Q}_{\kappa z} & =-{\cal Q}_{z\kappa}=\frac{\kappa}{\widetilde{\omega}}{R}_{ls}^{p}\,.
\end{align}
\end{subequations} }


%

\end{document}